\pdfoutput=1
\documentclass[11pt]{revtex4}
\usepackage[pdftex]{graphicx}
\usepackage{amsmath}
\usepackage{amssymb}
\usepackage{amsxtra}

\begin{document}

\title{
Can we predict the fourth family masses for quarks and leptons?\\
} 
\author{G. Bregar,  
N.S. Manko\v c Bor\v stnik \\
Department of Physics, FMF, University of Ljubljana, \\
Jadranska 19, SI-1000 Ljubljana, Slovenia}

\begin{abstract}
In the ref.
~\cite{norma,pikanorma,NF,NB}  four massless families of quarks and leptons before the 
electroweak break are predicted. Mass matrices of all the family members 
demonstrate in this proposal the same symmetry, determined by the family groups. 
There are scalar fields - two $SU(2)$ triplets, the gauge fields of the family quantum 
numbers, and three singlets, the gauge fields of the three  charges ( $Q, Q'$ and $Y'$)-
all doublets with respect to the weak charge,  which determine mass matrices on  
the tree level and, together with other contributions, also beyond the tree level. 
The symmetry of mass matrices remains unchanged for all loop corrections. 
The three singlets are, in loop corrections also together with other contributors, 
responsible for the differences in properties of the family members.   
Taking into account by the {\it spin-charge-family} theory proposed symmetry of mass 
matrices for all the family members and simplifying study by assuming that mass 
matrices are Hermitian and real and mixing matrices 
real, we fit free parameters of mass matrices to experimental data within the experimental 
accuracy. Calculations are in progress.
\end{abstract}
\maketitle

\section{Introduction}
\label{introduction}

There are several attempts in the literature to reconstruct mass matrices of quarks and leptons 
out of the observed masses and mixing matrices and correspondingly to learn more about properties 
of fermion families~\cite{FRI}. The most popular is the $n\times n$ matrix, 
close to the democratic one, predicting that $(n-1)$ families must be very light in 
comparison with the $n^{\rm th}$ one. 
Most of attempts treat neutrinos differently than the other family members, relying on the 
Majorana part, the Dirac part and the  "sea-saw" mechanism. Most often  are the number of families 
taken to be equal to the number of the so far observed families, while symmetries of mass matrices 
are chosen in several different ways~\cite{AstriAndOthers}. Also possibilities with four families 
are discussed~\cite{four}. 

In this paper we follow the prediction of the {\it spin-charge-family} 
theory~\cite{norma,pikanorma,NF,NB,GNBled2012} that there are four massless families above the 
electroweak break  and that the scalar fields - the two triplets carrying the family charges 
in the adjoint representations and the three singlets carrying the charges of  the 
family members ($Q,Q'$ and $Y'$) - all doublets with respect to the weak charge,   
cause (after getting nonzero vacuum expectation values) the electroweak break. Assuming that the 
contributions of all the scalar (and in loop corrections also of other) fields to mass matrices 
of fermions are real and symmetric, we are left with the following symmetry of mass matrices
 \begin{equation}
 \label{M0}
 {\cal M}^{\alpha} = \begin{pmatrix} 
 - a_1 - a & e     &   d & b\\ 
 e     & - a_2 - a &   b & d\\ 
 d     & b     & a_2 - a & e\\
 b     &  d    & e   & a_1 - a
 \end{pmatrix}^{\alpha}\,,
 \end{equation}
the same for all the family members $\alpha\in \{u,d,\nu,e \}$.  In appendix~\ref{M0SCFT} the 
evaluation of this mass matrix is presented and the symmetry commented. 
A change of  phases of the left handed and the right handed basis - there are ($2n-1$) free choices - 
manifests in a change of phases of mass matrices. 

The differences in the properties of the family members originate in the different charges of the family 
members and correspondingly in the different couplings to the corresponding scalar and gauge fields.

We fit (sect.~\ref{numericalresultsexp}) the mass matrix {Eq.~(\ref{M0})} with $6$ free parameters 
of any family member  $6$  
to the so far observed properties of quarks and leptons 
within the experimental accuracy. That is: {\it For a pair of either quarks or leptons, 
we fit twice $6$ free parameters of the two mass matrices  to twice three so far measured masses 
and to the corresponding mixing matrix}. Since we have the same number of free parameters (two times $6$ for 
each pair, since the mass matrices are assumed to be real) as there are measured quantities 
(two times $3$ masses and $6$ angles of the orthogonal mixing matrix under a simplification that 
the mixing matrix is real and Hermitian), we would predict the fourth family masses 
uniquely, provided that the measured quantities are accurate. The $n-1$ submatrix of any unitary 
matrix determine the unitary matrix uniquely for $n\ge 4$. The experimental inaccuracy enable to 
determine only the interval for the fourth family masses. 

If the prediction of the {\it spin-charge-family} theory, that there are four families which 
manifest in the massless basis the symmetry of Eq.~(\ref{M0}), is correct, we expect that enough 
accurate experimental data for the properties of the so far observed three families will offer 
narrow enough intervals for the fourth family masses.  

We treat all the family members, the quarks and the leptons, equivalently. We also estimate the 
contributions of the fourth family members to the mesons decays in dependence of 
the fourth family masses, 
taking into account also the estimations of the refs.~\cite{vysotsky}. However, we must admit that our 
estimations are so far pretty rough.

In sect.~\ref{numericalresultstoy} we check on a toy model how accurate must be the 
experimental data that enable the prediction of the fourth family masses:   
For two "known" mass matrices, obeying the symmetry of Eq.~(\ref{M0}), which lead approximately to the 
experimental data, we calculate masses  and the mixing matrix.  
Then, taking the mixing matrix and twice three lower masses as an input, we look back for the starting
two mass matrices with the required symmetry, allowing for the three lower families "experimental" 
inaccuracy. 
In the same section we then estimate the fourth family masses. So far the results are preliminary.
Although we spent quite a lot of efforts to make the results transparent and trustable, the 
numerical procedure to take into account the experimental inaccuracy of data is not yet good enough
to allow us to determine the interval of the fourth family masses, even not for quarks, so that all 
the results are very preliminary.

Still we can say that the so far obtained support the prediction of the {\it spin-charge-family} theory 
that there are four families of quarks and leptons, the mass matrices of which manifest the symmetry 
determined by the family groups -- the same for all the family members, quarks and leptons. 
The mass matrices are quite close to the "democratic" ones, in particular for leptons.

Since the mass matrices offer an insight into the properties of the scalar fields, which  determine 
mass matrices (together with other fields), manifesting effectively as the observed Higgs and the 
Yukawa couplings, we hope to learn about  the properties of these scalar fields also from the 
mass matrices of quarks and leptons.


In appendix~\ref{scft} we offer a very brief introduction into the {\it spin-charge-family} theory, 
which the reader, accepting the proposed symmetry of mass matrices without knowing  the 
origin of this symmetry, can skip. 

In sect.~\ref{procedure} the procedure to fit free parameters of mass matrices (Eq.~(\ref{M0})  
to the experimental data is discussed. 
We comment our studies in sect.~\ref{discussions}.

\section{Procedure used to fit free parameters of mass matrices to experimental data}
\label{procedure}
%


Matrices, following from the {\it spin-charge-family} theory might not be Hermitian 
(appendix~\ref{nonhermitean}). We, however, simplify our study, presented in this paper, by assuming 
that the mass matrix for any family member, that is for the quarks and the leptons, is real and symmetric. 
We take the simplest phases up to  signs, which depend on the choice of phases of the basic states, 
as discussed in appendices~\ref{M0SCFT}~\footnote{ 
In the ref.~\cite{gmdn} we made a similar assumption, except that we allow  that the  symmetry on the 
tree level of mass matrices might be changed in loop corrections. We got in that study dependance of mass matrices 
and correspondingly mixing matrices for quarks  on masses of the fourth family.}.

The matrix elements of mass matrices, with the loop corrections in all orders taken into account, 
manifesting the symmetry of Eq.~(\ref{M0}), are in this paper taken as free parameters.

Let us first briefly overview properties of mixing  matrices, a  more detailed explanation  of which 
can be found in subsection~\ref{extensionofmassm} of this section.

Let $M^{\alpha}$, $\alpha$ denotes the family member  ($\alpha=u,d,\nu,e$), be the mass matrix in the 
massless basis (with all loop corrections taken into account). 
Let  $V_{\alpha \beta}= S^{\alpha} S^{\beta \dagger}$,  where $\alpha$  
 represents either  the $u$-quark  and $\beta$ the $d$-quark, 
 or $\alpha$ represents the $\nu$-lepton and $\beta$ the $e$-lepton, denotes a (in general unitary)
 mixing 
 matrix of a particular pair.

 For $n\times n$ matrix ($n=4$ in our case) it follows: \\
 i. If a known submatrix  $(n-1) \times (n-1)$  of an unitary matrix $n \times n$ with $n\ge 4$ is extended 
 to the whole unitary  matrix  $n \times n$,  the $n^2$ unitarity conditions determine ($2(2(n-1) +1)$) real 
 unknowns completely.  
 If the submatrix $(n-1) \times (n-1)$ of an unitary matrix is made unitary by itself, then we loose 
 the information.\\
 ii. If the mixing matrix is assumed to be orthogonal, then the $(n-1) \times (n-1)$ submatrix contains 
 all the   information about the $n\times n $ orthogonal matrix to which it belongs and the $n(n+1)/2 $ 
 conditions  determine the $2(n-1)+1$ real unknowns completely for any $n$. \\
 If the submatrix of the orthogonal matrix is made orthogonal by itself, then we loose the 
 information.
 %

 We make in this paper, to simplify the present study, several assumptions~\cite{GNBled2012}, presented 
 already in the introduction. In what follows we present the procedure used in our study and repeat the 
 assumptions. 
 \begin{enumerate}
 \item If the mass matrix $ M^{\alpha}$ is Hermitian, then the unitary matrices  $S^{\alpha}$ and 
 $T^{\alpha}$, introduced in appendix~\ref{nonhermitean}  to diagonalize  a non Hermitian mass matrix, 
 differ only in phase factors depending on  phases  of basic vectors and manifesting in two diagonal matrices, 
 $F^{\alpha \,S}$ and  $ F^{\alpha\,T}$, 
 corresponding to  the left handed and the right handed basis, respectively. For Hermitian mass matrices we 
 therefore have:  $T^{\alpha}=S^{\alpha}\, F^{\alpha \,S} F^{\alpha\,T\, \dagger}$. By changing phases 
 of basic vectors we can change phases of $(2n-1)$ matrix elements.
 
 \item We take the diagonal matrices ${\cal M}^{\alpha}_{d}$ and the mixing matrices $V_{\alpha \beta}$ 
 from the available experimental data. 
 %
 The mass matrices $ M^{\alpha}$ in Eq.~(\ref{M0}) have, if they are Hermitian and real,  
 $6$ free real parameters  $(a^{\alpha}, a^{\alpha}_1, a^{\alpha}_2, b^{\alpha},  e^{\alpha},d^{\alpha})$.
%
 \item We limit the number of free parameters of the  mass matrix of each family member $\alpha$ 
 by taking into account $n$ relations among free parameters, in our case $n=4$, determined by the 
 invariants 
 \begin{eqnarray}
 \label{invariants}
I^{\alpha}_{1} &=& - \sum_{i=1,4} \,  m^{\alpha}_i, \;\quad
I^{\alpha}_{2} = \sum_{i>j=1,4} \,  m^{\alpha}_i \, m^{\alpha}_j,\;\nonumber\\
I^{\alpha}_{3} &=& -\sum_{i>j>k=1,4} \,  m^{\alpha}_i \, m^{\alpha}_j\, m^{\alpha}_k,\;\quad  
I^{\alpha}_{4} = m^{\alpha}_1\, m^{\alpha}_2 \,  m^{\alpha}_3\, m^{\alpha}_4\,, 
 \end{eqnarray}
which are expressions appearing at  powers of $\lambda_{\alpha}$,  $\lambda_{\alpha}^4+ $ 
$\lambda_{\alpha}^3 I_{1} +$ $\lambda_{\alpha}^{2} I_{2} + $ $\lambda_{\alpha}^1 I_{3} + $ 
$\lambda_{\alpha}^0 I_{4}=0$, in the eigenvalue equation.  
The invariants are fixed, within the experimental accuracy of the data,  by the observed masses of 
quarks and leptons and by the  fourth  family mass, if we make a choice of it. 
In appendix~\ref{reductionofpar}  we present the relations among the reduced number of free parameters 
for a  chosen  $m^{\alpha}_4$. There are ($6-4$) free parameters left   for each mass matrix.  
 \item
 The diagonalizing matrices $S^{\alpha}$ and $S^{\beta}$, each depending on the reduced number of free parameters,  
 are for real and symmetric mass matrices orthogonal. They follow from the 
 procedure 
 \begin{eqnarray}
 \label{mdiag}
 M^{\alpha}&=& S^{\alpha} \,{\bf M}^{\alpha}_{d}\,T^{\alpha \, \dagger}\,, \quad 
 T^{\alpha}=S^{\alpha}\, F^{\alpha \,S} F^{\alpha\,T\, \dagger}\,,\nonumber\\
 {\bf M}^{\alpha}_{d}&=& (m^{\alpha}_1, m^{\alpha}_2, m^{\alpha}_3, m^{\alpha}_4)\,,
 \end{eqnarray}
 provided that $S^{\alpha}$ and $S^{\beta}$ fit the experimentally observed mixing matrices 
 $V^{\dagger}_{\alpha \beta}$ within the experimental accuracy and that $ M^{\alpha}$  and $ M^{\beta}$ 
 manifest the symmetry presented in Eq.~(\ref{M0}).  We keep the symmetry of the mass matrices accurate. 
 One can proceed in two ways. 
 \begin{eqnarray}
 \label{SVud}
 A.:\;\;S^{\beta} = V^{\dagger}_{\alpha \beta} S^{\alpha}\,\,,&&
 \quad B.:\;\;S^{\alpha } = V_{\alpha \beta} S^{\beta}\,, \nonumber\\
 A.:\;\;V^{\dagger}_{\alpha \beta}\,S^{\alpha }\, {\bf M}^{\beta }_{d} \,S^{\alpha \dagger }
 V_{\alpha \beta}= M^{\beta} \,\,,&&\quad 
 B.:\;\;V_{\alpha \beta}\,S^{\beta }\, {\bf M}_{d}^{\alpha} \,S^{\beta \dagger } 
 V^{\dagger}_{\alpha \beta}= M^{\alpha} \,.
 \end{eqnarray}
 %
 In the  case $A.\,$  one obtains from Eq.~(\ref{mdiag}),  
 after requiring that the  mass matrix $M^{\alpha}$ has the desired  symmetry, 
 the matrix $S^{\alpha}$ and the mass matrix $M^{\alpha}$ ($=S^{\alpha } 
 {\bf M}^{\alpha}_{d} \,S^{\alpha \dagger }$), from where we get the mass matrix $M^{\beta}$ $= 
 V^{\dagger}_{\alpha \beta}\,S^{\alpha }\, {\bf M}^{\beta }_{d} \,S^{\alpha \dagger }
 V_{\alpha \beta}$. 
 In case  $B.\,$  one obtains  equivalently 
 the matrix $S^{\beta}$, from where we get  $M^{\alpha}$ ($=V_{\alpha \beta}\,S^{\beta }\, 
 {\bf M}_{d}^{\alpha} \,S^{\beta \dagger } V^{\dagger}_{\alpha \beta}$). 
 We use both ways iteratively taking into account the experimental accuracy of masses and mixing matrices. 
 \item Under the assumption of the  present study  that the mass matrices  
 are real and symmetric, 
 the orthogonal diagonalizing matrices $S^{\alpha}$ and $S^{\beta}$ form  the orthogonal mixing matrix 
 $V_{\alpha \beta}$, which depends on at most $6\,(=\frac{n(n-1)}{2})$  free real  parameters 
 (appendix~\ref{nonhermitean}).  
 Since, due to what we have explained at the beginning of this section, the  experimentally measured  
 matrix elements of the $3 \times 3$ submatrix of the $4 \times 4$  mixing matrix (if not made orthogonal 
 by itself)  determine the $4 \times 4$ mixing matrix - within the  experimental accuracy -  completely, 
 also the fourth family masses are determined, again within the experimental accuracy. 
 We must not forget, however, that the assumption of the real and symmetric mass matrices, leading to 
 orthogonal  mixing matrices, might not be an acceptable simplification, since we do know that the 
 $ 3\times 3$  submatrix of the mixing matrix has one complex phase, while the unitary $4\times 4$ 
 has three complex phases. (In the next step of study, with hopefully more accurate experimental data, 
 we shall relax  conditions on hermiticity of mass matrices and correspondingly on orthogonality of 
 mixing matrices.) 
 We expect that too large experimental inaccuracy  leave the fourth family masses in the present study 
 quite  undetermined, in particular for leptons.
\item We study quarks and leptons equivalently. The difference among family members originate on the 
tree level in the  eigenvalues of  the operators $(Q^{\alpha},Q'^{\alpha},Y'^{\alpha})$, which in loop 
corrections together with other contributors in  all orders contribute to all  mass matrix elements 
and cause the difference among family members~\footnote{There are also Majorana like terms  
contributing in  higher order loop corrections~\cite{NF} which might strongly influence in particular 
the neutrino mass matrix.}.
\end{enumerate}

Let us conclude. If the mass matrix of a family member obeys the symmetry required by the 
{\it spin-charge-family} theory, which in a simplified version (as it is taken in this study) is real and  
symmetric, the matrix elements of the  mixing matrices of quarks and leptons are correspondingly real, each 
of them with $\frac{n(n-1)}{2}$ free parameters.   
These six parameters of each mixing matrix are, within the experimental inaccuracy, determined by 
the three times three experimentally determined submatrix. After taking into account three so far measured 
masses of each family member, the six parameters of each mass matrix  reduce to three. 
Twice three free parameters are within the experimental accuracy correspondingly determined by
the $3 \times 3$ submatrix of the mixing matrix. The fourth family masses are correspondingly 
determined - within the experimental accuracy.

The assumption that the two $3\times 3$ mixing matrices are unitary would lead to the loss of the
information about the  $4\times 4 $ mixing matrix. 
This is the case also if we  take the orthogonalized version of the $3\times 3$ mixing matrices.

 Since neither the measured masses nor the measured mixing matrices are determined  accurately enough 
 to reproduce the $4 \times 4$ mixing matrices, 
 we can expect that the masses and mixing matrix elements of the fourth family will be determined  only 
 within  some quite large intervals.

\subsection{Submatrices and their extensions to unitary and orthogonal matrices}

\label{extensionofmassm}

In this appendix  well known properties of $n\times n$ matrices, extended from $(n-1)\times (n-1)$ 
submatrices are discussed. We make a short overview of the properties, needed in this paper, 
although all which will be presented here, is the knowledge  on the level of text books.

Any $n \times n$ complex matrix has $2n^2$ free parameters. The  $n+ 2 n(n-1)/2$ unitarity requirements 
reduce the number of free parameters to $n^2$ ($=2n^2-(n+ 2 n(n-1)/2)$). 

Let us assume a $(n-1) \times (n-1)$ known submatrix of the unitary matrix. The submatrix can be 
extended to the unitary matrix by ($2 \times [2(n-1)+1]$) real parameters of the last column and last line. 
The $n^2$ unitarity conditions on the whole matrix reduce the number of unknowns  to ($ 2(2n-1)  - n^2$). 
For $n=4$ and higher the $(n-1) \times (n-1)$ submatrix contains all the information about the unitary 
$n \times n$ matrix. 
The ref.~\cite{jarlskog} proposes a possible extension of an $(n-1) \times (n-1)$ unitary matrix 
$V_{(n-1)(n-1)}$ into  $n \times n$ unitary matrices $V_{n n}$.

The choice of phases of the left and the right basic states which determine the unitary matrix (like this 
is the case with the mixing matrices of quarks and leptons) reduces the number of free parameters 
for $(2n-1)$. Correspondingly is the number of free parameters of such an unitary matrix equal to $n^2- (2n-1)$,
which manifests in $\frac{1}{2}n(n-1)$ real parameters and $\frac{1}{2}(n-1)(n-2)$ 
$(=n^2-\frac{1}{2}n(n-1)-(2n-1) )$ phases (which determine the number of complex parameters).

Any real $n\times n$ matrix has $n^2$ free parameters which the $\frac{1}{2} n(n+1)$ orthogonality 
conditions reduce to  $\frac{1}{2} n(n-1)$. 
The $(n-1) \times (n-1)$ submatrix of this orthogonal matrix can be extended to this  
$n\times n$ orthogonal matrix with  $[2(n-1)+1]$ real parameters. The $\frac{1}{2} n(n+1)$ orthogonality 
conditions reduce these $[2(n-1)+1]$ free parameters to $(2n-1- \frac{1}{2} n(n+1))$,  which means that 
the $(n-1)\times (n-1)$ submatrix of an $n\times n$ orthogonal matrix determine 
properties of its $n\times n$ orthogonal matrix completely. Any $(n-1) \times (n-1)$ submatrix of 
an orthogonal matrix contains all the information about the whole matrix for any $n$.
Making the submatrix of the orthogonal matrix orthogonal by itself one looses the information about the 
$n \times n$ orthogonal matrix.

\subsection{Free parameters of mass matrices after taken into account invariants}
\label{reductionofpar} 

It is useful for numerical evaluation purposes to take into account for each family member its mass 
matrix invariants (sect.~\ref{invariants}), expressible with three within the experimental accuracy  
known masses, while we keep the fourth one as a free parameter. We shall make a choice of $a^{\alpha}$ 
instead of the fourth family mass.

We shall skip in this section the family member index $\alpha$ and introduce new parameters 
 as follows
\begin{eqnarray}
\label{deqrplusminus}
a, b\,, \quad f=d+e  \,,  \quad g=d-e\,, \quad q=\frac{a_1 + a_2}{\sqrt{2}}\,, 
\quad r=\frac{a_1 - a_2}{\sqrt{2}}\,.
\end{eqnarray}
After making a choice of $a \frac{I_{1}}{4}$, that is of the fourth family mass, four invariants 
of Eq.~(\ref{invariants}) reduce the number of free parameters to $2$.  The four invariants 
therefore relate six parameters leaving three of them, the $a$ included as a free parameter, undetermined.
There are for each pair of family members the measured mixing matrix elements, assumed in this paper 
to be orthogonal and correspondingly determined by six parameters, which then fixes these two times 
$3$ parameters. The (accurately enough) measured  $3\times 3$ submatrix of the (assumed to be orthogonal)
$4\times 4$ mixing matrix namely determines these $6$ parameters within the experimental accuracy. 

Using the starting relation among the invariants and introducing into them  new parameters 
($a,b,f,g,q,r$) from Eq.~(\ref{deqrplusminus}) we obtain
\begin{eqnarray}
\label{iprime}
a &=& \frac{I_{1}}{4}\,,\nonumber\\
I'_{2}&=& 
-I_{2} + 6a^2 - q^2 - r^2 -2b^2 = f^{2} + g^{2}\,, \nonumber\\
I'_{3}&=& - \frac{1}{2b} (I_{3}- 2a I_{2}+ 4 a^2) = f^{2} - g^{2}\,,\nonumber\\
I'_{4}&=& I_{4} - a I_{3}  +  a^2 I_{2}-3 a^4 \nonumber\\
&=& 
\frac{1}{4} (q^2-r^2)^2+ (q^2+r^2) b^2 + \frac{1}{2} (q^2-r^2)\cdot(\pm)\cdot[\pm]\, 2 g f
+ b^2 (f^2 +g^2) + \frac{1}{4} (2 g f)^2\,.
\end{eqnarray}
%

We eliminate, using  the first two equations,  the parameters $f$ and $g$, expressing them as functions 
of $I'_{2}$ and $I'_{3}$,
%
%
which depend, for a particular family member, on the three known masses, the parameter $a$  and the three 
parameters $r$, $q$ and $b$. 
We are left with the four free parameters ($a,b,q,r$) and the below relation among these parameters
\begin{eqnarray}
\label{qr}
&&\{- \frac{1}{2} (q^4 + r^4) + (-2 b^2 + \frac{1}{2} (-I_2 +6 a^2- 2 b^2)) (q^2+r^2)\nonumber\\
&+& (I'_{4}- \frac{1}{4} ( (-I_{2} +6a^2-2b^2)^2 + I'^2_{3}) + b^2 (-I_{2} +6a^2-2b^2))\}^2 \nonumber\\
&=& - \frac{1}{4} (q^2-r^2)^2 ( (-I_{2} +6a^2-2b^2 - (q^2 +r^2))^2 - I'^{2}_{3})\,, 
\end{eqnarray}
which reduces the number of free parameters to $3$. These $3$ free parameters must be determined, 
together with the corresponding three parameters  of the partner, from the measured mixing matrix.

We eliminate one of the $4$ free parameters in Eq.~(\ref{qr}) by solving the cubic equation for, let us 
make a choice, $q^2$
\begin{eqnarray}
\label{q2}
\alpha q^6 + \beta q^4 + \gamma q^2+ \delta =0\,.
\end{eqnarray}
Parameter ($\alpha,\beta,\gamma,\delta $) depend on the $3$  free remaining parameters ($a,b,r$) 
and the three, within experimental accuracy, known masses. 

To reduce the number of free parameters from the starting $6$ in Eq.~(\ref{M0}) to the $3$ left after 
taking into account invariants of each mass matrix, we look for the solution of Eq~(\ref{q2}) for all  
allowed values for ($a,b,r$). We make a choice for $a$ in the interval of $(a_{min},a_{max})$, 
determined by the requirement that $a$, which solves the equations,  is a real number. 
Allowing only real values for parameters $f$ and $g$ we end up with the equation
\begin{eqnarray}
\label{binterval}
-I_{2} + 6 a^2 -2 b^2 - (q^2 + r^2)  &>& |\frac{I_{3} + 8 a^3 -2 a I_{2}}{2b}|\,,
\end{eqnarray}
which determines the maximal positive $b$ for $q=0=r$ and 
also the minimal positive value for $b$.  For each value of the parameter $a$ the interval  
$(b_{min}, b_{max})$, as well as the interval $(r_{min}=0, r_{max})$, follow when taking into 
account experimental values for the three lower masses. 

%
\section{Numerical results}
\label{numericalresults}

Taking into account the assumptions and the procedure explained in sect.~\ref{procedure} and in the 
ref.~\cite{GNBled2012} we are looking for the $4 \times 4$ in this paper taken to be real and symmetric 
mass matrices for quarks and leptons, obeying the symmetry of Eq.~(\ref{M0}) and  manifesting 
properties -- masses and mixing matrices -- of the so far observed three families of quarks and leptons  
in agreement with the experimental limits for the appearance of the fourth family masses and 
mixing matrix elements to the lower three families, as presented in the 
refs.~\cite{data,vysotsky}. We also take into account our so far made rough estimations of  possible 
contributions of the fourth family members to the decay of mesons.  More detailed estimations are 
in progress.

We hope that we shall be able to learn from the mass matrices of quarks and leptons also about 
the properties of the scalar fields, which cause masses of quarks and leptons, manifesting 
effectively so far as the measured Higgs and Yukawa couplings.

First we test the predicting power of our model in dependence of the experimental inaccuracy of masses 
and mixing matrices on a toy model: Starting with two known mass matrices with the symmetry 
of Eq.~(\ref{M0}) we calculate masses and from the two diagonalizing matrices also the mixing matrix.
From the known masses and mixing matrix, for which we allow "experimental inaccuracy", we check how does 
the reproducibility of the two starting mass matrices depend on the "experimental inaccuracy" and how 
does the " experimental inaccuracy" influence the fourth family masses. 

Then we take the $3 \times 3$ measured mixing matrices for quarks and leptons and the measured masses,
all with the experimental inaccuracy. Taking into account that the  $3 \times 3$ submatrix of the unitary 
$4 \times 4$ matrix determines, if measured accurately enough, the $4 \times 4$ matrix, we look for the 
twice $4 \times 4$ mass matrices with the symmetry of Eq.~(\ref{M0}), and correspondingly for the fourth 
family masses, for quarks and leptons. 

When extending the two so far measured $3 \times 3$ submatrices of the $4 \times 4$ mixing matrices we 
try to take into account as many experimental data as possible.

\subsection{Checking on a toy model how much does the symmetry of mass matrices (Eq.~(\ref{M0})) limit 
the fourth family properties}
\label{numericalresultstoy}

We check in this subsection on a toy model the reproducibility of the starting two mass matrices  
from the known two times three lower masses (say $m_{u_i}, m_{d_i}\,,i=(1,2,3)$) and the $3 \times 3$  
submatrix  (say $(V_{ud})_{i,j}\,,i,j=(1,2,3)$) 
of the $4\times 4$ unitary mixing matrix in dependence of the inaccuracy allowed for 
$m_{u_i}, m_{d_i}\,,i=(1,2,3)$ and $(V_{ud})_{i,j}\,,i,j=(1,2,3)$.

We take the following two mass matrices, chosen so that they reproduce to high extent the measured 
properties of quarks (masses and mixing matrix) for some  experimentally acceptable values for the 
fourth family masses and  also the corresponding mixing matrix elements.
\begin{equation}
 \label{M0toyu}
 {\cal M}^{toy_{u}} = \begin{pmatrix} 
 220985. & 119365. & 120065. & 204610. \\
 119365. & 218355. & 204610. & 120065. \\
 120065. & 204610. & 192956. & 119365. \\
 204610. & 120065. & 119365. & 190325.
 \end{pmatrix}\,,
 {\cal M}^{toy_{d}} = \begin{pmatrix} 
175825. & 174262. & 174290. & 175709. \\
 174262. & 175839. & 175709. & 174290. \\
 174290. & 175709. & 175640. & 174262. \\
 175709. & 174290. & 174262. & 175654.
 \end{pmatrix}\,.  
\end{equation}

Diagonalizing these two mass matrices we  find the following twice four masses 
\begin{eqnarray}
\label{toymasses}
{\bf M}^{toy_{u}}_{d}/MeV/c^2= (1.3,620.,172000.,650000.)\,,\nonumber\\
{\bf M}^{toy_{d}}_{d}/MeV/c^2= (2.9,55.,2900.,700000.)\,,
\end{eqnarray}
and the mixing matrix
\begin{equation}
 \label{vudtoy}
 V_{toy_{ud}} = \begin{pmatrix}
  -0.97286 & -0.22946 & -0.02092 & 0.02134 \\
   0.23019 & -0.97205 & -0.04607 & -0.00287 \\
   0.00976 & 0.04965 & -0.99872 & -0.00045 \\
 0.02143 & 0.00213 & -0.00013 & 0.99977
\end{pmatrix}\,.
\end{equation}

In order to simulate experimental inaccuracies (intervals of values for  twice three lower 
masses and for the matrix elements of the $3 \times 3$  submatrix of the above unitary 
$4\times 4$ matrix) and test the influence of these inaccuracies on the fourth family masses, 
we change the fourth family  mass $m_{u_4}$ in the interval ($(300 - 1200)$) GeV and check the
accuracy with which the matrix elements of the $3\times 3$ submatrix of the $4 \times 4$ unitary 
matrix are reproduced. We measure the averaged inaccuracy in $\sigma$'s~\footnote{ 
We define $\sigma$ as the difference of the reproduced mixing matrix elements and the 
exact matrix elements, following from the starting two mass matrices.}.   We keep in 
Table~\ref{Table sigma} the $d_4$  mass equal to $700$ GeV.
 \begin{table}
 \begin{center}
 \begin{tabular}{|c|c|c|c|c|c|c|c|}
 \hline
 $m_{u_4}$/GeV             & 300 & 500 & 600 & 650 &700 & 800 &1200\\
 \hline
 "exp.$\,$ inacc"/$ \sigma$& 4.0 & 1.0 & 0.29& 0.0 &0.25& 0.66& 1.6\\
 \hline 
 \end{tabular}
 \end{center}
 \caption{\label{Table sigma} The average inaccuracy in $\sigma$ of the mixing matrix 
 elements of the $3 \times 3 $ submatrix of the unitary quark mixing matrix~(Eq.(\ref{vudtoy})) in 
 dependence of the  fourth family mass of the $m_{toy_{u_4}}$-quark. $m_{toy_{d_4}}$  mass is kept 
 equal to $700$ GeV.} 
 \end{table}

Let us add that the  accuracy, with which the $3 \times 3$ submatrix of the $4 \times 4$ mixing matrix 
is reproduced, depends much less on $m_{toy_{d_4}}$ than it does on $m_{toy_{u_4}}$ in this toy model 
case. 

We  use this experience when evaluating intervals, within which the fourth family masses appear 
when taking into account the inaccuracies of the experimental data.

%

%
\subsection{Numerical results for the observed  quarks and leptons with mass matrices obeying Eq.~(\ref{M0})}
\label{numericalresultsexp}

We take for the quark  and lepton masses the experimental values~\cite{data}, recalculated to the 
$Z$ boson mass scale. We take from~\cite{data} also the experimentally declared inaccuracies for 
the so far measured $3 \times 3$ mixing matrices, taken in our calculations as submatrices of 
the $4 \times 4$ unitary mixing matrices and pay attention on the experimentally allowed values 
for the fourth family masses and other limitations presented in refs.~\cite{four,vysotsky}~\footnote{
M.I.Vysotsky and A.Lenz comment in their very recent papers that the fourth family is excluded 
provided that one assumes the {\it standard model} with one scalar field (the Higgs) and extends 
the number of families from three to four while using loop corrections when evaluating the 
decay properties of the Higgs. We have, however, several scalar fields and first estimates show that
the fourth family quarks might have masses close to $1$ TeV.}. We also 
have started to make our own rough estimations for limitations which follow from the meson decays 
to which the fourth family members participate. Our estimations are in progress.

The numerical procedure, tested in the toy model and working well in this case, must still be adapted 
to take experimental inaccuracies into account in a way to be able to see which values within the 
experimentally allowed ones are the most trustable from the point of view of the symmetries of the 
$4 \times 4$ mass matrices predicted by the {\it spin-charge-family} theory. 

Although the accurate enough mixing matrices and masses of quarks and leptons are essential for the 
prediction of the fourth family members masses, we still hope that even with the present accuracy of 
the experimental data the intervals for the fourth family masses shall not be too large, in particular 
not for quarks, for which the data are much more accurate than for leptons.
Let us point out that from so far obtained results  we are not yet able to predict the fourth family 
mass intervals, which would be reliable enough.

We therefore present some preliminary results. Let us  point out that all the mass matrices manifest 
within a factor less then $2$ the "democratic" view. This is, as expected, more and more the case,
the higher might be the fourth family masses, and in particular is true for the leptons.

\begin{itemize}
\item For quarks we take~\cite{data}:
\begin{enumerate}
\item The quark mixing matrix~\cite{data} $V_{ud}= S^{u }\, S^{d\, \dagger}$ 
\begin{equation}
 \label{vud}
 |V_{ud}|= \begin{pmatrix}
   0.97425 \pm 0.00022    &  0.2252 \pm 0.0009    &  0.00415 \pm 0.00049    & |V_{u_1 d_4}|\\ 
   0.230   \pm 0.011      &  1.006  \pm 0.023     &  0.0409  \pm 0.0011     & |V_{u_2 d_4}|\\
   0.0084  \pm 0.0006     &  0.0429 \pm 0.0026    &  0.89    \pm 0.07       & |V_{u_3 d_4}|\\
   |V_{u_4 d_1}|    &  |V_{u_4 d_2}|    & |V_{u_4 d_3}|      & |V_{u_4 d_4}|
\end{pmatrix}\,,
\end{equation}
determining  for each assumed and experimentally allowed set of values for the mixing matrix elements 
of the $3 \times 3$ submatrix the corresponding  fourth family mixing matrix elements ($|V_{u_i d_4}|$ 
and $|V_{u_4 d_j}|$) from the unitarity condition for the $4 \times 4$ mixing matrix.
\item The masses of quarks are taken 
at the energy scale of $M_{Z}$, 
while we take the fourth family masses as  free parameters. We allow the values from $300$ GeV up to more 
than TeV to see the influence of the experimental inaccuracy on the fourth family masses.  
%
\begin{eqnarray}
\label{mumd}
 {\bf M}^{u}_{d}/{\rm MeV/c^2} &=& (1.27 + 0.50 - 0.42 , 619 \pm 84 , 171\,700. \pm 3\,000. , m^{u_4}> 
 335\, 000.)\,,\nonumber\\ 
 {\bf M}^{d}_{d}/{\rm MeV/c^2} &=& (2.90 +1.24 -1.19 ,  55 +16 -15 ,   2\,890. \pm   90. , m^{d_4}> 
 300\,000.)\,.
 \end{eqnarray}
 \end{enumerate}
\item For leptons we take~\cite{data}:
\begin{enumerate}
\item We evaluate $3 \times 3$ matrix elements from the data~\cite{data}
\begin{eqnarray}
\label{neutrinoex}
&&7.05 \cdot 10^{-17} \le \Delta({\bf m}_{21}/{\rm MeV/c^2})^2 \le  8.34 \cdot 10^{-17}\,,\nonumber\\
&&2.07 \cdot 10^{-15} \le \Delta({\bf m}_{(31),(32)}/{\rm MeV/c^2})^2 \le  2.75 \cdot 10^{-15}\,,\nonumber\\
&&0.25 \le \sin^2 \theta_{12} \le 0.37\,,\quad
0.36 \le \sin^2 \theta_{23} \le 0.67\,,\nonumber\\
&&\sin^2 \theta_{13} < 0.035 (0.056)\,,\quad
\sin^2 2 \theta_{13}= 0.098 \pm 0.013\,,\quad
\end{eqnarray}
which means that $\frac{\pi}{4}-\frac{\pi}{10} \le\theta_{23}\le \frac{\pi}{4}+\frac{\pi}{10} $,
$\frac{\pi}{5.4}-\frac{\pi}{10} \le\theta_{12}\le \frac{\pi}{4}+\frac{\pi}{10} $,
$\theta_{13}< \frac{\pi}{13}$.\\ 
This reflects in 
the lepton mixing matrix $V_{\nu e}= S^{\nu }\, S^{e\, \dagger}$
\begin{equation}
 \label{nue1}
 |V_{\nu e}|= \begin{pmatrix}
  0.8224    &  0.5200     &  0.1552         & |V_{\nu_1 e_4}|   \\
      0.3249     &  0.7239     &  0.6014    & |V_{\nu_2 e_4}|\\ 
      0.4455    &  0.4498     &  0.7704     & |V_{\nu_3 e_4}|\\ 
  |V_{\nu_4 e_1}|    & |V_{\nu_4 e_2}|   & |V_{\nu_4 e_3}| & |V_{\nu_4 e_4}|  
\end{pmatrix}\,,
\end{equation}
determining  for each assumed value for any mixing matrix element within the experimentally allowed  
inaccuracy the corresponding  fourth family mixing matrix elements ($|V_{\nu_i e_4}|$ and $|V_{\nu_4 e_j}|$) 
from the unitarity condition for the $4 \times 4$ mixing matrix.
\item The masses of leptons are taken from~\cite{data} 
while we take the fourth family masses as  free parameters, checking 
how much does the experimental inaccuracy influence a possible prediction for the fourth family 
leptons masses and how does this prediction agree with experimentally allowed 
values~\cite{data,vysotsky}  for the fourth family lepton masses.

\begin{eqnarray}
\label{mumd}
 &&{\bf M}^{\nu}_{d}/{\rm MeV/c^2} = ( 1 \cdot 10^{-9},\, 9\cdot 10^{-9},\,  
5 \cdot 10^{-8},\, m^{\nu_4}> 90\,000.) \,,\nonumber\\ 
 && {\bf M}^{e}_{d}/{\rm MeV/c^2} = (0.486\,570\,161 \pm 0.000\,000\, \,042,\,\nonumber\\
 && \;102.718\,135\,9 \pm 0.000\,009\,2, \, 1746.24 \pm 0.20, m^{e_4}> 102\,000\,
 ) 
 \,.
 \end{eqnarray}
\end{enumerate}
\end{itemize}

Following the procedure explained in sect.~\ref{procedure} we look for the mass matrices 
for the $u$-quarks and  the $d$-quarks  and the $\nu$-leptons and the $e$-leptons 
by requiring that the mass matrices reproduce experimental data while manifesting symmetry 
of Eq.~(\ref{M0}), predicted by the {\it spin-charge-family} theory.

We look for several properties of the obtained mass matrices:
{\bf i.} We test the influence of the experimentally declared inaccuracy of the $3 \times 3$ 
submatrices of the $4 \times 4$ mixing matrices and of the  twice $3$  measured masses on the 
prediction of the fourth family masses. 
{\bf ii.} We look for how could different choices for the masses of the fourth family members 
limit the inaccuracy of particular matrix elements of the mixing matrices or the inaccuracy of 
the three lower masses of family members. 
{\bf iii.} We test how close to  a democratic mass matrix 
are the obtained mass matrices in dependence of the fourth family masses.

The numerical procedure, used in this contribution, 
is designed  for  quarks and  leptons. 

In the two next subsections~\ref{quarks}, \ref{leptons} we present some preliminary results for $4 \times 4$ mass 
matrices as they follow from the {\it spin-charge-family} theory for quarks and leptons, respectively.


\subsubsection{Mass matrices for quarks}
\label{quarks}

Searching for mass matrices with the symmetries of Eq.~(\ref{M0}) to determine the interval for the fourth 
family quark masses in dependence of the values of the mixing matrix elements within the experimental 
inaccuracy, we have not yet found a trustable way to extract which experimental inaccuracies of  the 
mixing matrix elements should be taken more and which less "seriously". We also need to evaluate 
more accurately the experimental limitations for the fourth family masses, originating in decay properties 
of mesons and other experiments.
Although in the toy model case the "inaccuracy" of the matrix elements leads very clearly to 
the right fourth family masses, this is not the case when the experimental data for the $3\times 3$ mixing 
matrix elements are known within the accuracy from $0.02\%$ to $12\%$. The so far obtained results 
can not yet make the choice among less or more trustable experimental values: We can not yet make  more 
accurate choice for those data which have large experimental inaccuracies.
 
We are still trying to improve our the procedure of searching for the masses of the fourth family quarks.

Let us still present two cases to demonstrate how do quark mass matrices change with respect to the 
fourth family masses: The first two mass matrices lead to the fourth family masses $m_{u_{4}}=300$ GeV and 
$m_{d_{4}}=700$ GeV, while the second two lead
to the fourth family masses $m_{u_{4}}=1\,200$ GeV and 
$m_{d_{4}}=700$ GeV.

\begin{itemize}
\item
 \begin{equation}
 \label{mmud1}
 M^{u} = \begin{pmatrix} 
 402673. & 256848. & 267632. & 329419. \\
 256848. & 402393. & 329419. & 267632. \\
 267632. & 329419. & 283918. & 256848. \\
 329419. & 267632. & 256848. & 283638.
 \end{pmatrix}\,,
M^{d} = \begin{pmatrix} 
176784. & 174262. & 174524. & 175473. \\
 174262. & 176816. & 175473. & 174524. \\
 174524. & 175473. & 174663. & 174262. \\
 175473. & 174524. & 174262. & 174695.
\end{pmatrix}\,,  
\end{equation}
\begin{equation}
 \label{vud1}
 V_{ud}= \begin{pmatrix}
 0.97365 & 0.22296 & 0.00225 & -0.04782 \\
  0.22276 & -0.97412 & 0.03818 & -0.00444 \\
  0.01071 & -0.03671 & -0.99927 & -0.0001 \\
 0.04761 & 0.00634 & 0.00018 & 0.99885
 \end{pmatrix}\,.
\end{equation}
The corresponding masses are 
\begin{eqnarray}
\label{mmdud1}
 {\bf M}^{u}_{d}/{\rm MeV/c^2} &=& (1.29957 , 620.002 , 172\,000. ,   300\,000.)\,,\nonumber\\ 
 {\bf M}^{d}_{d}/{\rm MeV/c^2} &=& (2.88508 , 55.024 ,  2\,899.99 ,   700\,000.)\,. 
 \end{eqnarray}
 \item
  \begin{equation}
  \label{mmud2}
  M^{u} = \begin{pmatrix} 
   351427. & 256907. & 257179. & 342730. \\
   256907. & 342353. & 342730. & 257179. \\
   257179. & 342730. & 343958. & 256907. \\
   342730. & 257179. & 256907. & 334884. 
  \end{pmatrix}\,,
 M^{d} = \begin{pmatrix}
  175762. & 174263. & 174289. & 175708. \\
  174263. & 175581. & 175708. & 174289. \\
  174289. & 175708. & 175898. & 174263. \\
  175708. & 174289. & 174263. & 175717. 
 \end{pmatrix}\,,  
 \end{equation}
 \begin{equation}
  \label{vud2}
  V_{ud}= \begin{pmatrix}
  -0.9743 & 0.22521 & -0.00366 & 0.00383 \\
   0.22515 & 0.97325 & -0.04567 & 0.00299 \\
   -0.00672 & -0.04532 & -0.99895 & -0.00019 \\
 0.00305 & -0.00378 & -0.00004 & 0.99999
 \end{pmatrix}\,.
 \end{equation}
 The corresponding masses are
 \begin{eqnarray}
 \label{mmdud2}
  {\bf M}^{u}_{d}/{\rm MeV/c^2} &=& (1.29957 , 620.002 , 172\,000. ,   1\, 200\,000.)\,,\nonumber\\ 
  {\bf M}^{d}_{d}/{\rm MeV/c^2} &=& (2.88508 , 55.024 ,  2\,899.99 ,   700\,000.)\,. 
  \end{eqnarray}
 \end{itemize}

We notice:\\
{\bf i.}    In both cases the required symmetry, Eq.~(\ref{M0}), is (on purpose) kept  very accurate. 
      \\
{\bf ii.}   In both cases the mass matrices of quarks look quite close to the "democratic" matrix, in the 
second case  slightly more than in the first case.
      \\
{\bf iii.}  The mixing matrix elements are in the second case much closer (within the experimental values are
$V_{11}$, $V_{12}$, $V_{13}$ and $V_{32}$, almost within the experimental values are 
$V_{21}$, $V_{22}$ and $V_{33}$) to the experimentally allowed values, than in the first case (almost 
within the experimental allowed values are only $V_{21}$, $V_{22}$ and $V_{23}$). 

These results suggest that the fourth family masses $m_{u_4}= 1\, 200$ GeV and  $m_{d_4}=  700$ GeV 
are much more  trustable than $m_{u_4}= 300$ GeV and  $m_{d_4}=  700$ GeV.

\subsubsection{Mass matrices for leptons}
\label{leptons}

We present here results for leptons, manifesting properties of the lepton mass matrices.
These results are less informative than those for quarks, since the experimental results are 
for leptons mixing matrix much less accurate than in the case of quarks and also masses are 
known less accurately.

We have
\begin{itemize}
\item
\begin{equation}
  \label{mnue0}
  M^{\nu} = \begin{pmatrix} 
  14\,021.  & 14\,968.  &  14\,968. & -14\,021. \\ 
  14\,968.  & 15\,979.  &  15\,979. & -14\,968. \\ 
  14\,968.  & 15\,979.  &  15\,979  & -14\,968. \\ 
 -14\,021.  &-14\,968.  & -14\,968. &  14\,021.   
  \end{pmatrix}\,,\;
   M^{e} = \begin{pmatrix} 
  28\,933.  & 30\,057.  &  29\,762. & -27\,207. \\
  30\,057.  & 32\,009.  &  31\,958. & -29\,762. \\
  29\,762.  & 31\,958.  &  32\,009. & -30\,057. \\
 -27\,207.  &-29\,762.  & -30\,057. &  28\,933. 
  \end{pmatrix}\,, 
  \end{equation}
which leads to the mixing matrix $V_{\nu e}$ 
 \begin{equation}
  \label{vnue0}
  V_{\nu e1}= \begin{pmatrix}
    0.82363    &  0.54671    &  -0.15082     & 0.  \\ 
   -0.50263    &  0.58049    &  -0.64062     & 0.  \\
   -0.26268    &  0.60344    &   0.75290     & 0.  \\
    0.         &  0.         &   0.          & 0.
 \end{pmatrix}\,,
 \end{equation}
and the masses 
\begin{eqnarray}
\label{mmdnue0}
 {\bf M}^{\nu}_{d}/{\rm MeV/c^2} &=& (5\cdot 10^{-9}\,, 1\cdot 10^{-8}\,,
 4.9 \cdot 10^{-8}\,,60\, 000.)  \,,\nonumber\\ 
 {\bf M}^{e}_{d}/{\rm MeV/c^2} &=& (0.510999\,,105.658\,, 1\,776.82\, 120\,000)\,. 
 \end{eqnarray}
 We did not adapt lepton masses to $Z_{m}$ mass scale. Zeros ($0.$) for the matrix elements concerning 
 the fourth family members means that the values are less than
 $10^{-5}$. 
 \end{itemize}
  
 We notice:\\
 {\bf i.}    The required symmetry, Eq.~(\ref{M0}), is kept very accurate. 
       \\
 {\bf ii.}   The mass matrices of leptons are very close to the "democratic" matrix.
       \\
 {\bf iii.}  The mixing matrix elements among the first three and the fourth family members are very 
 small, what is due to our choice, since the matrix elements of the $3 \times 3$ submatrix of the 
 $4  \times 4$ unitary matrix, predicted by the {\it spin-charge-family} theory are very inaccurately 
 known.

\section{Discussions and conclusions}
\label{discussions}

One of the most important open questions in the elementary particle physics is: Where do the family originate?
Explaining the origin of families would answer the question about the number of families 
possibly observable at the low energy regime, about the origin of the scalar field(s) and Yukawa 
couplings and would also explain differences in the fermions properties - the differences in masses and 
mixing matrices among family members -- quarks and leptons.  

Assuming that the prediction of the {\it spin-charge-family} theory that there are four rather than so 
far observed  three coupled families, the mass matrices of which demonstrate in the massless basis the
$SU(2)\times SU(2)$ symmetry of Eq.~(\ref{M0}), the same for all the family members - the quarks and 
the leptons - we look in this paper for:\\
{\bf i.} The origin of differences in the properties of the family members - quarks and leptons. \\ 
{\bf ii.} The allowed intervals for the fourth family masses. \\
{\bf iii.} The matrix elements in the mixing matrices among the  fourth family members and the three 
already measured ones. \\

Our calculations presented here are preliminary and in progress.

Let us tell that there are two kinds of the scalar fields in the {\it spin-charge-family} theory, 
responsible for the masses and mixing matrices of quarks and leptons (and consequently also for 
the masses of the weak gauge fields): The ones which distinguish among the family members and the other ones
which distinguish among the families. The differences between quarks and leptons and between $u$ and $d$ 
quarks and between $\nu$ and $e$ leptons originate in the first kind of the scalar fields, which  
carry $Q, Q'$ (the two charges which, like in the {\it standard model}, originate in the weak and hyper 
charge) and $Y'$ (which originates in the hypercharge and in the fermion quantum number, similarly as in the 
$SO(10)$ models).

The existence of four coupled families seems almost unavoidable for the explanation of the properties 
of the neutrino families if all the family members should start from the massless basis in an 
equivalent way: The $4\times 4$ mass matrix, very close to a democratic one, offers three almost 
massless (in comparison with the observed quarks and charged leptons masses) families and a very 
massive one.

Taking the symmetry of, to simplify the calculations assumed to be real and symmetric, $4\times 4$ 
mass matrices, we determine  $6$ free parameters of any of the mass matrices 
by requiring that the mass matrices lead to the observed properties of quarks and leptons. In both 
cases the $2$ times three masses and the (in this simplified study) orthogonal  mixing matrix with 
$6$ parameters, determine the $2 \times 6$ parameters (as required by the {\it spin-charge-family}
theory) of the two mass matrices  within the experimental accuracy. 

{\it The same procedure is used to study either quarks or leptons.} Expected results are not only the mass 
matrices, but also the intervals within which masses of the fourth families should be observed and the 
corresponding mixing matrices.

We developed  a special procedure to extract the dependence of the fourth family masses on the experimental
inaccuracy of masses and mixing matrices. 
Our test of this procedure on a toy model, in which we first postulate two mass matrices (leading 
to masses and mixing matrices very close to those of quarks), calculate  the masses and 
the mixing matrix, and then from three lowest masses and the $3 \times 3$ sub matrix of the unitary 
$4 \times 4$ mixing matrix calculate back the starting mass matrices and the fourth family masses, 
showed that the procedure leads very accurately to the starting mass matrices. 

When we use the same procedure to extract the properties of the fourth family members from the 
experimental data within the experimental inaccuracies, the procedure was not selective enough 
to make useful predictions. We are improving the procedure to be able to extract the intervals 
of the fourth family masses in dependence of the  accuracy of particular data.
{\it Yet the preliminary  results presented here show, that the masses of the fourth family quarks 
with $m_{u_{4}} > 1 {\rm TeV}$ lead to the mixing matrix much closer to the experimental data 
than does $m_{u_{4}} \approx 300 {\rm GeV}$}.

Let us conclude this report by pointing out that even     if we shall not be able to limit the mass 
intervals for the fourth family members strongly enough to be predictive, yet the accurate enough 
data for the $3 \times 3$ submatrix of the unitary mass matrix will sooner or later determine the 
$4 \times 4$
unitary matrix so that the predictions will be accurate enough.

\appendix
\section{A brief presentation of the {\it spin-charge-family} theory}
 \label{scft}

We present in this section a very brief introduction into the {\it spin-charge family} 
theory~\cite{norma,pikanorma,NF,NB}. 
The reader can skip this appendix  taking by the {\it spin-charge family} theory required 
symmetry of mass matrices of Eq.~(\ref{M0}) as an input to the study of properties
of the $4\times 4$ mass matrices -- with the parameters which depend on charges of the family members -- and  
can come  to this part of the paper, if and when   would 
like  to learn  where do families and scalar fields possibly originate from. 

Let us start by directing attention of the reader to one of the most open questions in the 
elementary particle physics and cosmology: Why do we have families, where do they originate 
and correspondingly where do scalar fields, manifesting as Higgs and Yukawa couplings, originate?
The {\it spin-charge-family}  theory is offering a possible explanation for the origin of families and  
scalar fields, and in addition 
for the so far  observed charges and the corresponding gauge fields.

There are, namely, two (only two) kinds of the Clifford algebra objects: 
One kind, the Dirac $\gamma^a$, takes care of the spin in $d=(3+1)$, while the spin in  $d\ge 4$ 
(rather than the total angular momentum) manifests in $d=(3+1)$ in the low energy regime  as the charges. 
In this part the {\it spin-charge family} theory is like the Kaluza-Klein theory, unifying spin 
(in the low energy regime, otherwise the total angular momentum) 
and charges, and offering a possible answer to the question about 
the origin of the so far observed charges and correspondingly also about the so far observed gauge fields. 
The second kind of the Clifford algebra objects, forming the equivalent representations with respect 
to the Dirac kind, recognized by one of the authors (SNMB), is responsible for the appearance of families 
of fermions. 

There are correspondingly also two kinds of gauge fields, which appear to manifest in $d=(3+1)$ as the so far 
observed vector gauge fields (the number of - obviously non yet observed - gauge fields grows with the 
dimension) and as the scalar gauge fields. The scalar fields are responsible, after gaining nonzero 
vacuum expectation values, for the appearance of masses of fermions and gauge bosons. They manifest as
the so far observed Higgs~\cite{LHC} and the Yukawa couplings. 

All the properties of fermions and bosons in the low energy regime originate in the {\em spin-charge-family}
theory in a simple starting action for massless fields in $d=[1 +(d-1)]$. 
Fermions interact with the vielbeins $f^{\alpha}{}_a$ and correspondingly with the two kinds 
of the spin connection fields: with $\omega_{abc}= f^{\alpha}{}_{c}\,\omega_{ab \alpha}$ which are  
the gauge fields of $S^{ab}= \frac{i}{4}\,(\gamma^a \gamma^b - \gamma^b \gamma^a)$ and  with 
$\tilde{\omega}_{abc} =f^{\alpha}{}_{c}\, \tilde{\omega}_{ab \alpha}$ which are 
the gauge fields of $\tilde{S}^{ab}= \frac{i}{4}\, (\tilde{\gamma}^a \tilde{\gamma}^b - 
\tilde{\gamma}^b \tilde{\gamma}^a)$.  $\alpha, \beta,\dots$ is the Einstein index and 
$a,b,\dots$ is the flat index. The starting action is the simplest one
\begin{eqnarray}
\label{action}
S            \, = \int \; d^dx \; E\;{\mathcal L}_{f} &+&    
                \int \; d^dx \; E\; (\alpha \,R + \tilde{\alpha} \, \tilde{R})\,,\nonumber\\             
{\mathcal L}_f  = &&\frac{1}{2} (\bar{\psi} \, \gamma^a p_{0a} \psi) + h.c.\, \nonumber\\
p_{0a }         = f^{\alpha}{}_a \, p_{0\alpha} + \frac{1}{2E}\, \{ p_{\alpha}, E f^{\alpha}{}_a\}_- \,,
&& \,  
p_{0\alpha}     =  p_{\alpha}  -  \frac{1}{2}  S^{ab} \omega_{ab \alpha} - 
                    \frac{1}{2}  \tilde{S}^{ab}   \tilde{\omega}_{ab \alpha}\,,\nonumber\\                   \\ 
R               =  \frac{1}{2} \, \{ f^{\alpha [ a} f^{\beta b ]} \;(\omega_{a b \alpha, \beta} 
- \omega_{c a \alpha}\,\omega^{c}{}_{b \beta}) \} + h.c.\,,&&\,   
\tilde{R}       = \frac{1}{2}\,   f^{\alpha [ a} f^{\beta b ]} \;(\tilde{\omega}_{a b \alpha,\beta} - 
\tilde{\omega}_{c a \alpha} \tilde{\omega}^{c}{}_{b \beta}) + h.c.\,. 
\end{eqnarray}
Fermions, coupled to the vielbeins and the two kinds of the spin connection fields, {\it  manifest} (after 
several breaks of the starting symmetries) {\it before the electroweak break four massless families 
of quarks and leptons}, the left handed fermions are weak charged and the right handed ones are weak 
chargeless. 
The vielbeins and the two kinds of the spin connection fields manifest effectively as the observed 
gauge fields and (those with the scalar indices in $d=(1+3)$) as  several scalar fields.  The mass 
matrices of the four family members (quarks and leptons) are after the electroweak break expressible 
on a tree level by the vacuum expectation values of the two kinds of the spin connection fields and the 
corresponding vielbeins with the scalar indices (\cite{NB,NBLED2}):\\
{\bf i.} One kind originates in the scalar fields $\tilde{\omega}_{abc}\,$, manifesting   
as  the two $SU(2)$ triplets -- $\tilde{A}_{s}^{\tilde{N}_L \,i}, i=(1,2,3)\,,s=(7,8)$; 
$\tilde{A}^{\tilde{1}\,i}_{s}\,, i=(1,2,3)\,,s=(7,8)$; --  and one singlet -- $\tilde{A}_{s}^{\tilde{4}}\,, 
s=(7,8)$ -- contributing equally to all the family members. \\
{\bf ii.} The second kind originates in the  scalar fields $\omega_{abc}$, manifesting as 
 three singlets  -- $A^{Q}_{s}, A^{Q'}_s, A^{Y'}\,, s=(7,8)$ -- contributing the same values to all 
 the families and distinguishing among family members. $Q$ and $Q'$ are the quantum numbers  from 
 the {\it standard model}, $Y'$ originates in the second $SU(2)$ (a kind of a right handed "weak") charge.

All the scalar fields manifest, transforming the right handed quarks and leptons into the 
corresponding  left handed ones~\footnote{It is the term $\gamma^0 \gamma^s \,\phi^{Ai}_s\,$, 
where $\phi^{Ai}_s$, with $s=(7,8)$ denotes any of the scalar fields, which transforms the right handed 
fermions into the corresponding left handed partner~\cite{NF,NB,NBLED2}. This mass term originates in 
$\bar{\psi} \, \gamma^a p_{0a} \psi$ of the action~Eq.(\ref{action}), with $a= s=(7,8)$ and $p_{0s}= 
f^{\sigma}_{s}\, (p_{\sigma} - \frac{1}{2} \tilde{S}^{ab}\tilde{\omega}_{ab \sigma} - 
\frac{1}{2} S^{st}\omega_{st \sigma})$.}
and contributing also to the masses of the weak bosons,  as doublets with respect to the 
weak charge.
Loop corrections, to which all the scalar and also gauge vector fields contribute coherently, change 
contributions of the off-diagonal and diagonal elements on the tree level, keeping the tree level symmetry 
of mass matrices unchanged~\footnote{It can be seen that all the loop corrections keep the starting symmetry 
of the mass matrices unchanged. We have also started~\cite{NF,AN} with the evaluation of  the loop 
corrections to the 
tree level values.  
This estimation has been done so far~\cite{AN} only up to the  first order and partly to the second order.}.

\subsection{Mass matrices on the tree level and beyond which manifest $SU(2)\times SU(2)$ symmetry}
\label{M0SCFT}

Let us make a choice of a massless basis $\psi_{i}$, $i=(1,2,3,4)$, for a particular family memeber 
$\alpha$. And let us take into account the two kinds of the operators, which transform the basis 
vectors into one another
\begin{eqnarray}
\label{taunl}
\tilde{N}^{i}_{L}\,,\,i=(1,2,3)\,, \quad \quad \tilde{\tau}^{i}_{L}\,,\,i=(1,2,3)\,,
\end{eqnarray}
with the properties
\begin{eqnarray}
\label{taunlonpsi}
&&\tilde{N}^{3}_{L}\, (\psi_1, \psi_2,\psi_3,\psi_4)= \frac{1}{2} (-\psi_1, \psi_2,-\psi_3,\psi_4)\,,\nonumber\\
&&\tilde{N}^{+}_{L}\, (\psi_1, \psi_2,\psi_3,\psi_4)=  (\psi_2, \;\;0,\psi_4,\;\;0)\,,\nonumber\\
&&\tilde{N}^{-}_{L}\, (\psi_1, \psi_2,\psi_3,\psi_4)=  (0\;\;, \psi_1,\;\;0,\psi_3)\,,\nonumber\\
&&\tilde{\tau}^{3}\, (\psi_1, \psi_2,\psi_3,\psi_4)= \frac{1}{2} (-\psi_1, -\psi_2,\psi_3,\psi_4)\,, \nonumber\\
&&\tilde{\tau}^{+}\, (\psi_1, \psi_2,\psi_3,\psi_4)= (\psi_3, \psi_4,\;\;0,\;\;0)\,,\,,\nonumber\\ 
&&\tilde{\tau}^{-}\, (\psi_1, \psi_2,\psi_3,\psi_4)= (\;\;0, \;\;0,\psi_1,\psi_2) \,.
\end{eqnarray}
This is indeed what the two $SU(2)$ operators in the {\it spin-charge-family} theory do. The gauge 
scalar fields of these operators determine, together with the corresponding coupling constants, the off diagonal 
and diagonal matrix elements on the tree level. In addition to these two kinds of $SU(2)$ scalars there are 
three $U(1)$ scalars, which distinguish among the family members, contributing on the tree level the same
diagonal matrix elements for all 
the families. In loop corrections in all orders the symmetry of mass matrices remains unchanged, while the
three $U(1)$ scalars,  contributing coherently with the two kinds of $SU(2)$ scalars  and all the massive 
fields to all the matrix elements, manifest in off diagonal elements as well.  
  All the scalars are doublets with respect to the weak charge,  
contributing to the weak and the hypercharge of the fermions so that they transform the right handed members into 
the left handed onces.

With the above (Eq.~(\ref{taunlonpsi}) presented choices of phases of the left and the right handed basic 
states in the massless basis the mass matrices of all 
the family members manifest the symmetry, presented in Eq.~(\ref{M0}). 
One easily checks that a change of the phases of the left and the right handed members, there are $(2n-1)$ 
possibilities, causes  changes in phases of matrix elements in Eq.~(\ref{M0}).

\section{Properties of non Hermitian mass matrices}
\label{nonhermitean}

This pedagogic presentation of well known properties of non Hermitian matrices can be 
found in many textbooks, for example~\cite{chine}. We repeat this topic here only to make our discussions  
transparent.

Let us take   a  non Hermitian mass matrix  $M^{\alpha}$ as it follows from the 
{\it spin-charge-family} theory, $\alpha$ denotes a family member 
(index ${}_{\pm}$ used in the main text is dropped).

We always can diagonalize  a non Hermitian $M^{\alpha}$ with two unitary matrices, 
$S^{\alpha}$ ($S^{\alpha\, \dagger}\,S^{\alpha}=I$)
and $T^{\alpha}$ ($T^{\alpha\, \dagger}\,T^{\alpha}=I$)
\begin{eqnarray}
\label{diagnonher0}
S^{\alpha\, \dagger}\,M^{\alpha}\,T^{\alpha}&=& {\bf M}^{\alpha}_{d}\,=
(m^{\alpha}_{1}\, \dots m^{\alpha}_{i}\, \dots m^{\alpha}_{n}).
\end{eqnarray}
The proof is added below. 

Changing phases of the basic states, those of the left handed one and those of the right handed one, the 
new unitary matrices  $S'^{\alpha} = S^{\alpha} \,F_{\alpha S} $ and $T'^{\alpha} = T^{\alpha}\,F_{\alpha T}$
change the phase of the elements of diagonalized mass matrices ${\bf M}^{\alpha}_{d}$ 
\begin{eqnarray}
\label{diagnonher}
S'^{\alpha\, \dagger}\,M^{\alpha}\,T'^{\alpha}&=& F^{\dagger}_{\alpha S}\,{\bf M}^{\alpha}_{d}\,
F_{\alpha T}=\nonumber\\
& & diag(m^{\alpha}_{1} e^{i(\phi^{\alpha S}_{1}- \phi^{\alpha T}_{1})}\, \dots 
m^{\alpha}_{i}\, e^{i(\phi^{\alpha S}_{i}- \phi^{\alpha T}_{i})}\,, \dots m^{\alpha}_{n}\,
e^{i(\phi^{\alpha S}_{n}- \phi^{\alpha T}_{n})})\,,\nonumber\\
F_{\alpha S} &=& diag(e^{-i \phi^{\alpha S}_{1}},\,\dots\,,e^{-i \phi^{\alpha S}_{i}}\,,\dots\,,
e^{-i \phi^{\alpha S}_{n}})\,,\nonumber\\
F_{\alpha T} &=& diag(e^{-i \phi^{\alpha T}_{1}},\,\dots\,,
e^{-i \phi^{\alpha T}_{i}}\,,\dots\,, e^{-i \phi^{\alpha T}_{n}})\,.
\end{eqnarray}

In the case that the mass matrix is Hermitian $T^{\alpha}$ can be replaced by $ S^{\alpha}$, but only up 
to phases originating in the phases of the two basis, the left handed one and the right handed one, 
since they remain independent. 

One can diagonalize the non Hermitian mass matrices in two ways, that is either one diagonalizes  
$M^{\alpha }M^{\alpha \,\dagger}$ or $M^{\alpha \dagger} M^{\alpha }$
\begin{eqnarray}
\label{diagMM}
(S^{\alpha \dagger} M^{\alpha} T^{\alpha})          (S^{\alpha \dagger} M^{\alpha } T^{\alpha})^{\dagger}&=&
S^{\alpha \dagger} M^{\alpha } M^{\alpha \,\dagger} S^{\alpha} = {\bf M}^{\alpha  2}_{d S}\,, \nonumber\\  
(S^{\alpha \dagger} M^{\alpha} T^{\alpha})^{\dagger}(S^{\alpha \dagger} M^{\alpha } T^{\alpha})&=&
T^{\alpha \dagger} M^{\alpha\, \dagger} M^{\alpha } T^{\alpha} = {\bf M}^{\alpha  2}_{d T}\,, \nonumber\\
{\bf M}^{\alpha\, \dagger }_{d S}&=& {\bf M}^{\alpha }_{d S}\, , 
\quad {\bf M}^{\alpha\, \dagger }_{d T}= {\bf M}^{\alpha}_{d T}\,.
\end{eqnarray}
One can prove that ${\bf M}^{\alpha }_{d S}={\bf M}^{\alpha }_{d T}$. The proof proceeds as follows.
Let us define two Hermitian ($H^{\alpha}_{S}\,$, 
$H^{\alpha}_{T}$) and two unitary matrices ($U^{\alpha}_{S}\,$, $H^{\alpha}_{T}$) 
\begin{eqnarray}
\label{proof1}
H^{\alpha}_{S} &=& S^{\alpha} {\bf M}^{\alpha }_{d S} S^{\alpha \, \dagger}\,, \quad \quad
H^{\alpha}_{T}  =  T^{\alpha} {\bf M}^{\alpha \dagger }_{d T} T^{\alpha \, \dagger}\,,\nonumber\\
U^{\alpha}_{S} &=& H^{\alpha -1}_{S} M^{\alpha } \,, \quad \quad U^{\alpha}_{T} = H^{\alpha -1}_{T} 
M^{\alpha \,\dagger } \,,
\end{eqnarray}
It is easy to show  that $H^{\alpha\, \dagger}_{S}= H^{\alpha}_{S}$, 
$H^{\alpha\, \dagger}_{T}= H^{\alpha}_{T}$, $U^{\alpha}_{S} \,U^{\alpha\, \dagger}_{S}=I$ and 
$U^{\alpha}_{T} \,U^{\alpha\, \dagger}_{T}=I$.
Then it follows  
\begin{eqnarray}
\label{proof2}
S^{\alpha \dagger}\, H^{\alpha}_{S} \,S^{\alpha} &=& {\bf M}^{\alpha }_{d S}= {\bf M}^{\alpha \,\dagger}_{d S}=
S^{\alpha \dagger}\,M^{\alpha }\,U^{\alpha \,-1}_{S} \,S^{\alpha}= S^{\alpha \dagger}\,M^{\alpha }\, T^{\alpha} \,,
\nonumber\\
T^{\alpha \dagger}\, H^{\alpha}_{T} \,T^{\alpha} &=& {\bf M}^{\alpha }_{d T}= {\bf M}^{\alpha \,\dagger}_{d T}=
T^{\alpha \dagger}\,M^{\alpha \, \dagger}\,U^{\alpha \,-1}_{T} \,T^{\alpha}= T^{\alpha \dagger}\,
M^{\alpha \dagger}\, S^{\alpha} \,,
\end{eqnarray}
where we recognized $U^{\alpha \,-1}_{S} \,S^{\alpha}= T^{\alpha}$ and $U^{\alpha \,-1}_{T} \,T^{\alpha}=S^{\alpha}$. 
Taking into account Eq.~(\ref{diagnonher}) the starting basis can be chosen so, that all diagonal 
masses are real and positive.

\end{document}